\documentclass{osa-article}

\journal{osac}

\articletype{Research Article}

\newcommand{\RomanNumeralCaps}[1]
    {\MakeUppercase{\romannumeral #1}}

\begin{document}

\title{Tomographic phase and attenuation extraction for a sample composed of unknown materials using X-ray propagation-based phase-contrast imaging}

\author{S.~J.~Alloo,\authormark{1*} D.~M.~Paganin, \authormark{2} K.~S.~Morgan, \authormark{2} T.~E.~Gureyev, \authormark{3,2,4,5} S.~C.~Mayo, \authormark{6} S.~Mohammadi, \authormark{7,8} D.~Lockie, \authormark{9} R.~H.~Menk, \authormark{7} F.~Arfelli, \authormark{10} F.~Zanconati, \authormark{11} G.~Tromba, \authormark{7} K.~M.~Pavlov \authormark{1,2,5}}

\address{\authormark{1}School of Physical and Chemical Sciences, University of Canterbury, Christchurch, New Zealand\\
\authormark{2}School of Physics and Astronomy, Monash University, Victoria, Australia\\
\authormark{3}School of Physics, University of Melbourne, Parkville, Victoria, Australia\\
\authormark{4}Faculty of Health Sciences, University of Sydney, Lidcombe, Australia\\
\authormark{5}School of Science and Technology, University of New England, Armidale, Australia\\
\authormark{6}Commonwealth Scientific and Industrial Research Organisation, Melbourne, Australia\\
\authormark{7}Elettra-Sincrotrone Trieste, Trieste, Italy\\
\authormark{8}The Abdus Salam ICTP, Trieste, Italy\\
\authormark{9}Maroondah BreastScreen, Melbourne, Australia\\
\authormark{10}Department of Physics, University of Trieste and INFN Trieste, Italy\\
\authormark{11}Department of Medical Science--Unit of Pathology, University of Trieste, Trieste, Italy}

\email{\authormark{*}samantha.alloo@pg.canterbury.ac.nz} 

\begin{figure}[htb]
    \centering
    \includegraphics[width=0.9\linewidth]{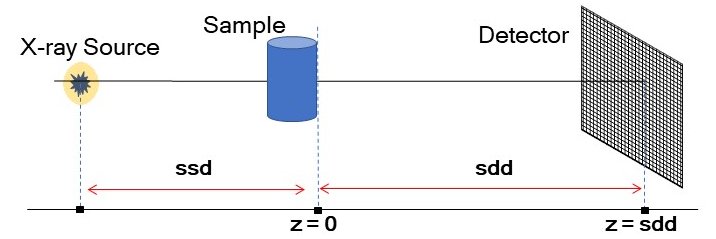}
    \caption{\textit{Schematic of experimental set-up for propagation-based phase-contrast X-ray imaging.}}
    \label{fig:PBPCI}
\end{figure}

\begin{figure}[htb]
    \centering
    \includegraphics[width=0.8\textwidth]{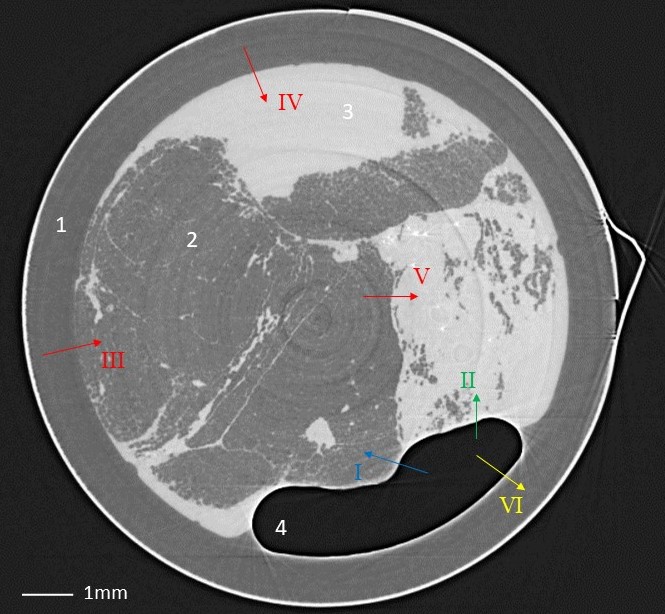}
    \caption{\textit{PB-PCXI CT of a breast-tissue sample. Composite materials, polypropylene, adipose, glandular tissue, and air are labeled 1, 2, 3, and 4, respectively. \RomanNumeralCaps{1} - \RomanNumeralCaps{6} denote line profiles taken across various interfaces in the sample.}}
    \label{fig:tissue}
\end{figure}

\begin{abstract*}
Propagation-based phase-contrast X-ray imaging (PB-PCXI) generates image contrast by utilizing sample-imposed phase-shifts. This has proven useful when imaging weakly-attenuating samples, as conventional attenuation-based imaging does not always provide adequate contrast. We present a PB-PCXI algorithm capable of extracting the X-ray attenuation, \textit{$\beta$}, and refraction, \textit{$\delta$}, components of the complex refractive index of distinct materials within an unknown sample.
The method involves curve-fitting an error-function-based model to a phase-retrieved interface in a PB-PCXI tomographic reconstruction, which is obtained when Paganin-type phase-retrieval is applied with incorrect values of $\delta$ and $\beta$. The fit parameters can then be used to calculate true \textit{$\delta$} and \textit{$\beta$} values for composite materials. This approach requires no \textit{a priori} sample information, making it broadly applicable. Our PB-PCXI reconstruction is single distance, requiring only one exposure per tomographic angle, which is important for radiosensitive samples. We apply this approach to a breast-tissue sample, recovering the refraction component, \textit{$\delta$}, with 0.6 - 2.4\% accuracy compared to theoretical values.\\\\
\end{abstract*}
\noindent Attenuation-based X-ray radiography relies on absorption and scatter of X-rays traversing a material. In attenuation regimes, the registered intensity images are proportional to the negative exponential of the object's projected linear attenuation coefficient, $\mu(\textbf{r})$, along straight-line ray paths \cite{Kak_1988}. 
Attenuation-based techniques can image objects whose projected attenuation varies significantly over the detector plane, but this approach is insufficient when this variation is small. Phase-contrast X-ray imaging (PCXI) \cite{PCGoetz,Ingal_1995, pfeiffer_weitkamp_bunk_david_2006, Frster1980DoubleCD, Bonse_1965,PBSingirev,PBWilkins,PBNugent, Cloetens1996,Olivo2001} is a non-destructive imaging method that has proven particularly useful in imaging weakly-attenuating samples. PCXI techniques, including grating-based \cite{pfeiffer_weitkamp_bunk_david_2006,Momose}, analyzer-based \cite{Frster1980DoubleCD,Ingal_1995,PCGoetz}, interferometric \cite{Bonse_1965}, edge-illumination \cite{Olivo2001,Olivo2021} and propagation-based (PB-PCXI) \cite{PBSingirev,PBWilkins,PBNugent} approaches, consider refraction effects, described by $\delta(\textbf{r})$, as well as attenuation, described by $\beta(\textbf{r})$, where \textit{$n(\textbf{r})=1-\delta(\textbf{r})+i\beta(\textbf{r})$} is the complex refractive index, as a function of position \textit{$\textbf{r}$}. 
\\\\PB-PCXI, achieved using the set-up in Fig.~\ref{fig:PBPCI}, visualizes phase-contrast effects via Fresnel diffraction fringes \cite{PBSingirev,Cloetens1996} formed during free-space propagation of transmitted X-rays. PB-PCXI phase-retrieval algorithms are often employed to obtain projected phase, attenuation and/or thickness information from the detector measured intensity. Paganin \textit{et al}.~\cite{TIEHom} derived a noise-robust deterministic phase-retrieval method for PB-PCXI, for the case of a single-material object. This algorithm requires \textit{a priori} sample knowledge via an input parameter  $\gamma = \delta/\beta$. The approach in Ref.~\cite{TIEHom} is single-distance, which becomes important when imaging radiosensitive samples, 
as radiation dose can be diminished. Such phase-retrieval algorithms have also proven to increase the signal-to-noise ratio \cite{Beltran:10, Beltran_2018, Gureyev:17}.
\\\\
Paganin {\em et al.}'s~\cite{TIEHom} phase-retrieval algorithm has been extended to allow for multi-material objects \cite{Beltran:10} and partially-coherent sources \cite{Beltran_2018}. Beltran \textit{et al.}~\cite{Beltran:10} reported a computed tomography (CT) PB-PCXI algorithm capable of correctly phase-retrieving pairs of adjacent materials within a multi-material object. However, this algorithm requires \textit{a priori} knowledge of the complex refractive index for each material present in the sample, limiting its application when exact sample composition is unknown. Thompson \textit{et al.}~\cite{Thompson_2019} used the homogeneous form of the transport of intensity equation \cite{Teague}, in a similar way to Paganin {\em et al.}~\cite{TIEHom}, to derive a three-dimensional phase-retrieval algorithm for PB-PCXI CT data. In this letter, we extend these two and three-dimensional algorithms~\cite{TIEHom, Beltran:10, Beltran_2018, Thompson_2019} to the case of multi-material objects, aiming to independently extract refractive and absorption properties without \textit{a priori} sample knowledge. The proposed method 
may be viewed as a deterministic multi-material extension of the iterative single-material method for electron microscopy described in Eastwood \textit{et al.}~\cite{Eastwood:11}. 
\\\\ We begin with Eqn.~18 from Thompson \textit{et al.}~\cite{Thompson_2019}, which describes the three-dimensional distribution of the $\delta$ component of a single-material object's complex refractive index, which can be transformed to the $\beta$ component since $\gamma=\delta/\beta$ is constant: 
\begin{equation}
    \beta_{\textrm{Recon.}}(x,y,z) = (1/2k)\left[1-\tau\nabla^2\right]^{-1}\Re \mathfrak{F}_{2}K_{\theta}(x,y,z).
    \label{eqn:ThompsonBeta}
\end{equation}
Above, $\Re$ is the filtered back-projection (FBP) operator \cite{Natterer}, $\mathfrak{F}_{2}$ is the two-dimensional Fourier transform, ${K_{\theta}(x,y,z)}$ is the in-line contrast function at sample angular orientation $\theta$ \cite{Thompson_2019}, and  $\nabla^2 = \partial^2/\partial x^2 + \partial^2/\partial y^2 + \partial^2/\partial z^2$ is the Laplacian. $\tau$ is related to the phase-retrieval input parameter, $\gamma$, for a single-material object, via $\tau=sdd\,\lambda\gamma/M4\pi,$ where $sdd$ is the  sample-to-detector propagation distance, $\lambda=2\pi/k$ is the X-ray wavelength, and $M=1+sdd/ssd$ is the sample magnification due to divergent X-rays, where \textit{ssd} is the source-to-sample distance.\\\\
\begin{figure}[htb]
    \centering
    \includegraphics[width=0.8\linewidth]{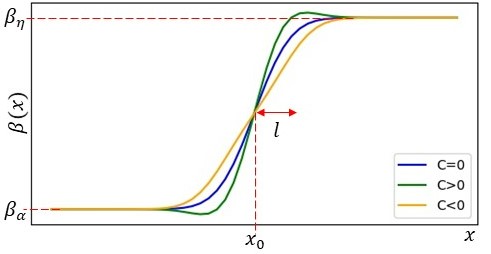}
    \caption{\textit{Line profiles, modeled using the right side of Eqn.~\ref{eqn:FullErrorFunctionModel}, demonstrating (green) under-smoothing and (yellow) over-smoothing effects of phase-retrieval, in algorithms implemented within CT reconstruction. The blue trace demonstrates correct phase-retrieval.}}
    \label{fig:errorextend}
\end{figure}
\\\\\\\\ Equation \ref{eqn:ThompsonBeta} can be applied to  model a profile of the reconstructed $\beta_{\textrm{Recon.}}$ across an interface between two materials, here denoted as materials $\alpha$ and $\eta$, by making the replacement of the phase retrieval parameter $\tau$ with $\tau_{\textrm{edge}}$, where we now define $\gamma_{\textrm{edge}}$ as \cite{Gureyev2002}: 
\begin{equation}
    \gamma_{\textrm{edge}} = \left[\delta_{\alpha}-\delta_{\eta}\right]/\left[\beta_{\alpha}-\beta_{\eta}\right].
    \label{eqn: GammaInterface}
\end{equation}
Furthermore, we consider Eqn.~\ref{eqn:ThompsonBeta}~in the case where $\gamma_{\textrm{edge}}$ is selected incorrectly for the given pair of interfaces within a multi-material object. We denote the correct input parameter by $\gamma_{\textrm{edge}}$ and the incorrect parameter by $\gamma_{\textrm{edge}}'$, and follow the same convention for $\tau_{\textrm{edge}}$. The value for $\gamma_{\textrm{edge}}'$ will result in under- or over-smoothed interfaces in the reconstructed CT image. To consider these effects, we follow Beltran \textit{et al.}~\cite{Beltran:10}, and apply the operator $[2k(1-\tau_{\textrm{edge}}\nabla^2)/2k(1-\tau_{\textrm{edge}}'\nabla^2)]$ to both sides of Eqn.~\ref{eqn:ThompsonBeta}. This operator describes the non-step-like behavior seen at material interfaces when $\gamma_{\textrm{edge}}$ is selected incorrectly, with $\gamma_{\textrm{edge}}'$. Applying this operator, and retaining terms of only first order in $\nabla^2$ in the Taylor series expansion of the left-hand side, gives:
\begin{align}
    \left[1+(\tau_{\textrm{edge}}' - \tau_{\textrm{edge}})\nabla^2\right]\beta_{\textrm{True}}(x,y,z) = \nonumber\\{\left(1/2k\right)\left[1-\tau_{\textrm{edge}}'\nabla^2\right]^{-1}} \Re \mathfrak{F}_{2} K_{\theta}(x,y,z).
    \label{eqn:RevisedBeta}
\end{align}
The right-hand side of this expression represents the reconstructed three-dimensional distribution of the attenuation coefficient, $\beta_{\textrm{Recon.}}(x,y,z)$,  for an incorrect $\tau_{\textrm{edge}}'$. 
\\\\To proceed, we consider Eqn.~\ref{eqn:RevisedBeta} in one transverse direction, \textit{x}, such that two materials $\alpha$ and $\eta$ are spanned. Under this consideration, the correct reconstructed attenuation coefficient, $\beta_{\textrm{True}}(x,y,z)$, in Eqn.~\ref{eqn:RevisedBeta}, that is with no over- or under-smoothing effects, can be modeled by an error-function, given by the general form: 
\begin{equation}
    \beta_{\textrm{True}}(x) = \frac{\beta_{\alpha}+\beta_{\eta}}{2}+\frac{\beta_{\eta}-\beta_{\alpha}}{2}~ {\textrm{erf}}\left(\frac{x-x_o}{l}\right).
\label{eqn:errorfunct_basic}
\end{equation}
Here, $\beta_{\alpha}$ and $\beta_{\eta}$ are the uniform $\beta$ values that are taken on either side of the interface (outside of the PB fringe), $l$ is the interface width, \textit{x} is the position coordinate in a direction perpendicular to the interface located at $x=x_o$, and $\textrm{erf}(x)$ represents an error-function, as defined in Eqn.~7.1.1 of Abramowitz and Stegun~\cite{ErrorEqn}. The error-function comes from the convolution of a step function (sharp interface) with a Gaussian. This Gaussian can describe either the imaging system point-spread function (PSF) \cite{Gureyev_2004} and/or an interface that is not perfectly sharp, due to mixing of the two materials at the interface. The blue curve in Fig.~\ref{fig:errorextend} plots Eqn.~\ref{eqn:errorfunct_basic}, describing a profile across an interface within a phase-retrieved CT reconstruction, for the case where $\gamma_{edge}$ is chosen correctly for the pair of materials making up that interface. 
\\\\
Substituting Eqn.~\ref{eqn:errorfunct_basic} into the left-hand side of Eqn.~\ref{eqn:RevisedBeta} takes us to a relationship between the incorrect $\tau_{\textrm{edge}}'$ and true value, $\tau_{\textrm{edge}}$, for a given phase retrieved CT line profile, $\beta_{\textrm{Recon.}}(x)$,
\begin{gather}
    \beta_{\textrm{Recon.}}(x) = \frac{\left(\beta_{\alpha}+\beta_{\eta}\right)}{2} + \frac{\left(\beta_{\eta}-\beta_{\alpha}\right)}{2}\:{\textrm{erf}}\left(\frac{x-x_o}{l}\right)+\nonumber
    \\C\left(\frac{x-x_o}{l}\right)\:{\textrm{exp}}\left(-\frac{{\left(x-x_o\right)}^2}{l^2}\right),
    \label{eqn:FullErrorFunctionModel}
\end{gather}
where the coefficient \textit{C} is derived to be:  
\begin{equation}
    C = \frac{4(\beta_{\eta}-\beta_{\alpha})(\tau_{\textrm{edge}}-\tau_{\textrm{edge}}')}{2l^2\sqrt{\pi}}.
    \label{eqn:C}
\end{equation}
Equation \ref{eqn:FullErrorFunctionModel} can model the residual edge-enhancement (under-smoothing) and over-smoothing effects across an interface that is produced by an incorrect choice for $\gamma_{\textrm{edge}}$ \cite{Gureyev_2004,Beltran:10}. The green curve in Fig.~\ref{fig:errorextend} demonstrates how a positive value of \textit{C} in Eqn.~\ref{eqn:FullErrorFunctionModel} models residual edge-enhancement at the boundary of two materials. In the contrary case, the orange curve in Fig.~\ref{fig:errorextend} demonstrates the effect of over-smoothing, with a negative coefficient \textit{C}. Equations~\ref{eqn:FullErrorFunctionModel} and~\ref{eqn:C} can be used, in conjunction with curve-fitting techniques, to (i) determine the correct $\gamma_{\textrm{edge}}$ for a given boundary in a multi-material sample, and then (ii) reconstruct $\delta$ and $\beta$ for composite materials. The latter task can be achieved via a set of linear equations, with one equation per class of sample interface in the form of Eqn.~\ref{eqn: GammaInterface}, which can then be uniquely solved for $\delta$ for each composite material in the object. $\beta_{\alpha}-\beta_{\eta}$ in Eqn.~\ref{eqn: GammaInterface} can be directly measured from reconstructed CT slices, as variations of $\gamma_{\textrm{edge}}$ do not affect reconstructed $\beta$ values far away from the given interface \cite{Gureyev:13}. To uniquely solve the system of linear equations, and extract $\delta$ for all composite materials in the sample, the following criteria should be met: (i) The number of unique interfaces in the sample has to be greater than, or equal to, the number of composite materials; (ii) One reference material, for which $\delta$ is known, is required. The reference material can be vacuum, where $\delta=0$.  Usually the sample is surrounded by either air ($\delta_{\textrm{air}}\approx0$ ) or a known material, so this is not an onerous requirement.
\\\\Our algorithm was applied to CT of a breast-tissue sample, shown in Fig.~\ref{fig:tissue}; this is the same dataset as labeled `Tissue 5c' in Gureyev \textit{et al.} \cite{Gureyev2014}. The tissue was inside a polypropylene tube, material \textit{1} in Fig.~\ref{fig:tissue}. The experimental CT data were collected at the Synchrotron Radiation for Medical Physics (SYRMEP) ELETTRA Beamline. A 20 keV quasi-monochromatic X-ray beam illuminated the sample, which was fixed on a rotation stage, with $ssd=23$ m and $sdd=1$ m. The detector was a water-cooled CCD camera (Photonic Science model VHR), $4008 \times 2672$ pixels full-frame, used in $2 \times 2$ binning mode (resulting in a pixel size of 9 $\mu$m), coupled to a gadolinium oxysulfide scintillator placed on a fiber optic taper.\\
\begin{figure}[htb]
    \centering
    \includegraphics[width=0.8\linewidth]{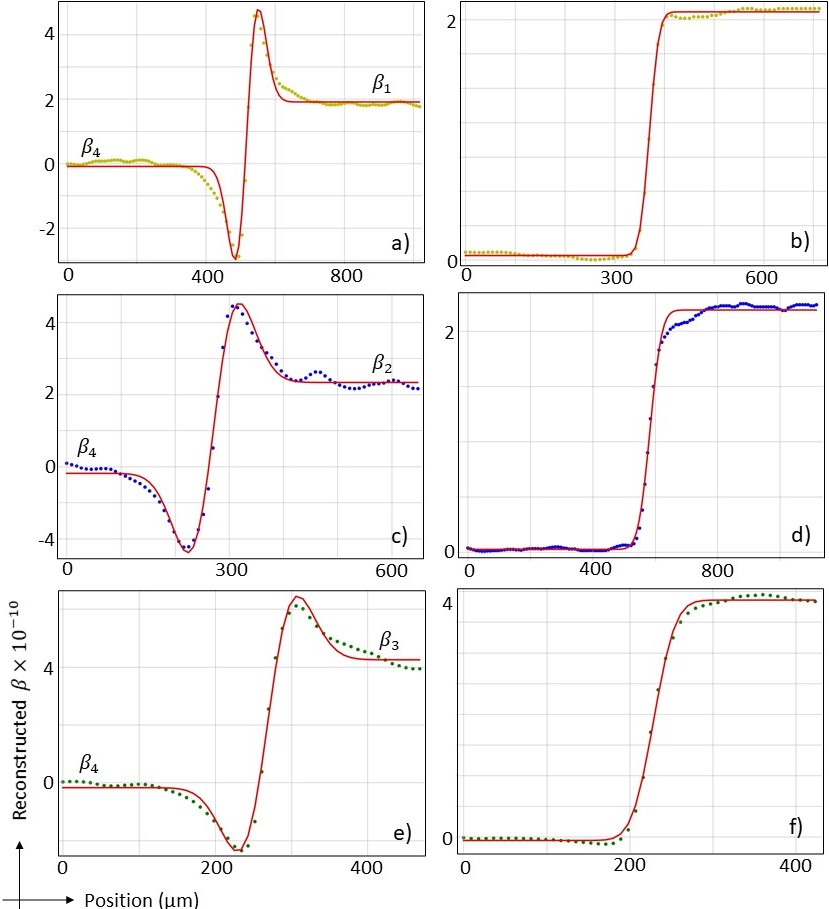}
    \caption{\textit{Line profiles across interfaces in the breast-tissue CT. Referring to Fig.~\ref{fig:tissue}, (top) line profile \RomanNumeralCaps{6}, (middle) line profile \RomanNumeralCaps{1}, and (bottom) line profile \RomanNumeralCaps{2}. (left) are taken from the incorrectly phase-retrieved CT image with $\gamma_{\textrm{edge}} = 350$, and (right) are taken when the correct $\gamma_{\textrm{edge}}$ for the given interface was used: b) $\gamma_{\textrm{edge}} = 2900$, d) $\gamma_{\textrm{edge}} = 2500$ and f) $\gamma_{\textrm{edge}} = 1430$.}}
    \label{fig:lineprofiles}
\end{figure} 

\noindent CT reconstructions, employing Hamming filtered back-projection, were performed using the XTRACT \cite{XTRAXT} implementation of Paganin \textit{et al.}'s single-material phase-retrieval algorithm \cite{TIEHom}, using $\gamma_{\textrm{edge}}=350$. One  reconstructed CT axial slice is shown in Fig.~\ref{fig:tissue}. Six line profiles, labelled \textit{\RomanNumeralCaps{1} - \RomanNumeralCaps{6}} in Fig.~\ref{fig:tissue}, were drawn across unique interfaces in the phase-retrieved CT slice. This initial choice of $\gamma_{\textrm{edge}}= 350$ correctly reconstructed interfaces \textit{\RomanNumeralCaps{4}}, and \textit{\RomanNumeralCaps{5}}, however residual edge-enhancement was seen across \textit{\RomanNumeralCaps{1}}, \textit{\RomanNumeralCaps{2}, \RomanNumeralCaps{3},} and \textit{\RomanNumeralCaps{6}}. The figures on the left of Fig.~\ref{fig:lineprofiles} show raw and fitted line profiles, \textit{\RomanNumeralCaps{1}}, \textit{\RomanNumeralCaps{2}}, and \textit{\RomanNumeralCaps{6}}, taken between air, labeled \textit{4} in Fig.~\ref{fig:tissue}, and composite materials, labeled \textit{1, 2,} and \textit{3}, in the breast-tissue sample. Curve-fits to Eqn.~\ref{eqn:FullErrorFunctionModel} were performed using a Levenberg-Marquardt algorithm \cite{press}, and the fit coefficients were extracted. These fit data were then used to calculate the correct $\gamma_{\textrm{edge}}$ for each interface, giving: $\gamma_{\textrm{edge}:\RomanNumeralCaps{1}} = 2500\pm100$, $\gamma_{\textrm{edge}:\RomanNumeralCaps{2}} = 1430\pm90$, $\gamma_{\textrm{edge}:\RomanNumeralCaps{3}} = 2000\pm1000$, $\gamma_{\textrm{edge}:\RomanNumeralCaps{4}} = 350\pm20$, $\gamma_{\textrm{edge}:\RomanNumeralCaps{5}} = 350\pm20$, and $\gamma_{\textrm{edge}:\RomanNumeralCaps{6}}=2900\pm200$. Here, the uncertainties were calculated using propagation of the one-standard-deviation errors of the curve-fit coefficients. CT reconstructions using each of these $\gamma_{\textrm{edge}}$ input parameters were performed, where the corresponding $\beta$ for the optimized materials, either side of the interface, could be measured. Instances of Eqn.~\ref{eqn: GammaInterface} for each interface in the sample established a set of linear equations which could be uniquely solved. In our case, the resultant system of linear equations was over-determined, hence QR factorization was used to give a least-squares solution \cite{press} for the refractive-index decrement, $\delta$, for composite materials in the breast-tissue. In these calculations $\delta_{4} \approx 0$ and $\beta_{4} \approx 0$, since material \textit{4} is known to be air, satisfying criterion (ii). 
\begin{table}[htbp]
\centering
\caption{\bf Coefficients of the index of refraction of composite materials (1 = polypropylene, 2 = adipose, 3 = gland) of the breast-tissue sample: 20keV X-rays}
\begin{tabular}{cccc}
\hline
 & 1 & 2 & 3\\
\hline
Calculated $\delta (\times10^{7})$ & $5.0\pm0.3$ & $5.4\pm0.3$ & $5.8\pm0.4$ \\
Theoretical $\delta (\times10^{7})$ &$5.03$ & $5.36$ & $5.94$\\
$\delta$: \% Difference & 0.60\% & 0.75\%  & 2.4\%\\
Calculated $\beta (\times10^{10})$ & $1.77\pm0.04$ & $2.17\pm0.04$ & $3.9\pm0.1$\\
Theoretical $\beta (\times10^{10})$ & $1.82$ & $2.54$ & $3.96$\\
$\beta$: \% Difference & 2.8\% & 15\% & 1.5\% \\
\hline
\end{tabular}
  \label{table}
\end{table}
\\\\Table \ref{table} shows the calculated, and theoretical \cite{Gureyev2014,Brennan}, components of the index of refraction for composite materials in the breast-tissue. Our approach determined the refractive-index decrement, $\delta$, to, at worst, 2.4\% accuracy. The small discrepancies are thought to be due to small intrinsic differences typically seen in identical biological samples. Note, the effects of residual phase-contrast were utilized in this analysis, i.e.~edge-enhancement at boundaries that remains after the phase retrieval has been performed. While our model in principle can admit negative \textit{C} values, that is, model over-smoothed interfaces, the reconstruction proposed here is more robust in a regime with under-smoothed interfaces, seen also in Eastwood \textit{et al.}'s electron microscopy phase-retrieval algorithm \cite{Eastwood:11}.
\\
\\In summary, we obtained refraction and attenuation information from X-ray phase contrast images a weakly-attenuating multi-material sample, breast-tissue, given no \textit{a priori} sample information. The method is more robust in the case when residual phase-contrast is seen as a result of phase-retrieval, that is boundaries are under-smoothed, hence the initial CT reconstruction should be performed with a sufficiently small choice of phase-retrieval input parameter, $\gamma$. The method has the potential to uniquely determine composite materials within an unknown sample. This may find application in various fields, including medicine, biology, paleontology, earth sciences, biosecurity, and multiple engineering disciplines such as failure prediction. The approach can be extended to situations where image-blurring effects, due to finite source size, are significant, such as a laboratory-source setting. This may be done by the replacement $\gamma\rightarrow{}\gamma-(2S^2/sdd)$ \cite{Beltran_2018}, where \textit{S} is the radius of the effective incoherent PSF at the detector plane.
\\\\\\
\textbf{ACKNOWLEDGMENTS}
This research was undertaken at the Synchrotron Radiation for Medical Physics (SYRMEP) ELETTRA Beamline. We acknowledge travel funding provided by the International Synchrotron Access Program (ISAP) managed by the Australian Synchrotron and funded by the Australian Government. Dr Morgan was supported by FT180100374. We acknowledge the University of Canterbury for awarding a Doctoral Scholarship to S.~J.~Alloo, and useful discussions with M.~J.~Kitchen and Ya.~I.~Nesterets. 
\bibliography{refs}
\end{document}